\documentclass{laynagenerictemplate}

\begin{document}

\maketitle \thispagestyle{empty}

\begin{abstract}
Capture-recapture methods for estimating the total size of elusive populations are widely-used, however, due to the choice of estimator impacting upon the results and conclusions made, the question of performance of each estimator is raised. Motivated by an application of the estimators which allow covariate information to meta-analytic data focused on the prevalence rate of completed suicide after bariatric surgery, where studies with no completed suicides did not occur, this paper explores the performance of the estimators through use of a simulation study. The simulation study addresses the performance of the Horvitz-Thompson, generalised Chao and generalised Zelterman estimators, in addition to performance of the analytical approach to variance computation. Given that the estimators vary in their dependence on distributional assumptions, additional simulations are utilised to address the question of the impact outliers have on performance and inference.
\end{abstract}


\Addresses
\section{Introduction and background}

Developed for use in ecology, capture-recapture methods are utilised for estimating the total size of elusive populations. An incomplete list of the individuals is used for the estimation, as given the nature of these populations, many individuals remain unobserved. Specifically for animal populations within ecology, traps are placed in a designated study area to capture the animals, where those in the captured sample are uniquely marked and released. On further occasions, additional samples of the animals are taken, recording previously marked individuals and uniquely marking the unmarked individuals. A capture history for each of the individual animals observed at least once used to estimate the total number of individuals in the population is achieved by repeating this process a predetermined number of times.

Capture-recapture methods have evolved to have utility in other fields, including those of the application of this paper, medicine and epidemiology. The focus of this paper is on meta-analytic data with missing zeroes motivated by a systematic review by \cite{peterhansel2013risk} which estimates the prevalence rate of completed suicide after bariatric surgery and where studies with a count of zero do not occur due to of the search criteria. In this setting, the independent studies can be treated as the individual animals, and the observational period for each of the studies is comparable to the trapping occasions. As the counts are zero-truncated, the number of studies with zero counts remains unknown and hence of interest to be estimated. Table~\ref{tab:datatable} in the appendix contains the data from the 27 observed studies in \cite{peterhansel2013risk}, including the number of completed suicides, person-years, proportion of women and the country of origin of each study. 

To estimate the total number of studies, and consequently the number of missing studies, a choice of which estimation method to use is required. However, the capture-recapture estimators approach data differently, resulting in estimates which can differ significantly from one another, impacting the accuracy and reliability of conclusions made. 

Motivated by the case study, the aim of this paper is to utilise a simulation study to compare the performance of capture-recapture estimators which allow for covariate information. For the performance analysis, uncertainty quantification is required, also assessing the performance of the conditional approach to the analytical variance computation. Analysis of the performance of both the capture-recapture estimators and of the variance formula are novel contributions, the conclusions of which can lead to more accurate population size estimates with reliable confidence intervals.

\section{Motivating application}\label{sec:methods}

 In summary, capture-recapture is the methodology of estimating population sizes when some individuals within the population go unobserved. The case study data contains the counts of completed suicide, $y_i$, 
 with corresponding person-years, $e_i$, for each study $i=1,\dots,n$, where $n=27$. For the random variable $Y_i$, it is assumed that $Y_1,\dots,Y_n$ are independent. Given that the data also include covariates, let $\mathbf{x}=(x_{i1}, x_{i2})^T$ be the vector denoting the covariates for the $i$th study, where $x_{i1}$ is the proportion of women of study $i$ and $x_{i2}$ is the country of origin for study $i$, given as
 \begin{equation*}
     x_{i2}=
     \begin{cases}
        1 & {\text{ if country origin is USA}},\\
        0 & {\text{ if otherwise}}.
     \end{cases}
 \end{equation*}

 Study 24. Smith 2004 does not include a value for the proportion of women, so is imputed as $x_{24,1}=0.823$ using a linear regression imputation model, where the model is chosen from the full model with backwards stepwise Bayesian information criterion (BIC), resulting in the proportion of women as the response variable and both person-years and country of origin as main effects with their interaction. Additionally, for model fitting purposes, the country of origin of study 21. Kral 1993 is changed from "USA/Sweden" to "USA" given that "USA" is both listed first, and is the country of origin with highest frequency of occurrence. 
 
 Whilst many approaches can be taken within capture-recapture, this paper focuses on the methods of the Horvitz-Thompson, generalised Chao and generalised Zelterman estimators \cite[see][]{horvitz1952generalization,bohning2013generalization,bohning2009covariate}. These methods utilise an expected count value computed from a regression model, and as a result, have the benefit of allowing for the inclusion of covariate information. Several assumptions are required in order to use these estimators. Firstly, the population is assumed to be a closed system, so there are no births or deaths. This assumption is often hard to meet, for example, if the data is focused on a disease with high mortality. However, since we are treating the studies as the individuals, rather than the study participants, this assumption of a closed population is met if no studies are added or removed from the systematic review. The second assumption is of independence between individuals, in this case, independence between the studies. Given that the studies are required to be independent from one another to be in the systematic review, it is reasonable to assume this is also met. Lastly, there is the assumption of independence between captures, in context meaning that within each study, the completed suicide of one individual does not affect the completed suicide of anybody else in the study. It is fair to assume that this true for the case study data, therefore, all assumptions required for capture-recapture are met for the case study data. 

\section{Population size estimators}\label{sec:estimation}
\subsection{Horvitz-Thompson estimator}\label{sec:HorThom}

Proposed by \cite{horvitz1952generalization}, the Horvitz-Thompson estimator is a popular capture-recapture population size estimator (see \cite{borchers2002estimating}, Chapter 11 and \cite{mccrea2014analysis}, Chapter 3). For a given regression function, $\mathbf{h}$  (see Table~\ref{tab:lp} in Appendix~\ref{app:tables} for more information), with corresponding coefficients, $\boldsymbol{\beta}$, the total number of studies, N, is given by
\begin{equation}
    \widehat{N}^{(HT)}=\sum_{i=1}^N \dfrac{I_i}{1-P(Y=0|\hat{\mu}_i)},
    \label{equ:nhte}
\end{equation}
where $\hat{\mu}_i=e_i\exp\left[\mathbf{h}(\mathbf{x}_i)^T\hat{\beta}\right]$ is the expected count of study $i$ and $I_i$ is an indicator variable defined as
\begin{equation*}
    I_i=
    \begin{cases}
        1 & \text{ if study $i$ is observed}, \\
        0 & \text{ otherwise}.
    \end{cases}
\end{equation*}
In our context, for $y=0,1,\dots$, the probability in (\ref{equ:nhte}) arises from either a Poisson model
\begin{equation*}
    P(Y=y)=\exp(-\mu)\frac{\mu^y}{y!},
\end{equation*}
or negative-binomial model
\begin{equation*}
    P(Y=y)=\frac{\Gamma(y+\alpha)}{\Gamma(y+1)\Gamma(\alpha)}\left(\frac{\alpha}{\mu+\alpha}\right)^\alpha\left(\frac{\mu}{\mu+\alpha}\right)^y.
\end{equation*}

For assessing uncertainty, the analytical variance can be computed using the conditional approach proposed by \citet[][page 314]{van2003point}, where the theoretical formula is as follows. 
\begin{equation}
    \widehat{\text{Var}}(\widehat{N}^{(HT)})=E[ \text{Var}(\widehat{N}^{(HT)}|I_i)]+\text{Var}(E[\widehat{N}^{(HT)}|I_i]).
    \label{equ:condvar}
\end{equation}

An approximation of the analytical variance is then given as 
\begin{equation}
        \widehat{\text{Var}}(\widehat{N}^{(HT)}) = \left(\sum_{i=1}^n \nabla G(\hat{\mu}_i|\hat{\boldsymbol{\beta}})\right)^T\text{Cov}(\hat{\boldsymbol{\beta}})\left(\sum_{i=1}^n \nabla G(\hat{\mu}_i|\hat{\boldsymbol{\beta}})\right) + \sum_{i=1}^n \frac{\exp(-\hat{\mu}_i)}{(1-\exp(-\hat{\mu}_i))^2},
        \label{equ:varht}
\end{equation}
where
\begin{equation*}
    \nabla G(\hat{\mu}_i|\boldsymbol{\hat{\beta}})=-\dfrac{\exp(\log(\hat{\mu}_i)-\hat{\mu}_i)}{(1-\exp(-\hat{\mu}_i))^2}\times \mathbf{h}(\mathbf{x}_i)^T.
\end{equation*}

Whilst widely-used, the Horvitz-Thompson estimator relies heavily on the entire data following the given distributional assumption. As a result, if the counts do not strictly follow the distribution, for example, if the data contain outliers as is often the case for real life data, the accuracy of the resulting population size estimate and precision of confidence intervals can be negatively affected. In addition, larger count values are more susceptible to deviation from the given distribution. Therefore, it is beneficial to explore alternative population size estimators which do not experience these issues and are more resilient to outliers.

\subsection{Generalised Chao estimator}\label{sec:genChao}

Developed as a method that approaches unobserved heterogeneity and focuses on estimating the lower bound of the population size, Chao's Lower Bound \citep{chao1987estimating} can be used as an alternative to the Horvitz-Thompson estimator. However, the conventional estimator of $N^{(C)}=n + \dfrac{f_1^2}{2f_2}$, where $f_y$ is the frequency of exactly $Y=y$ counts, does not allow for covariate information, and hence does not allow for the incorporation of an exposure variable either. Adapted by \citet{bohning2013generalization} to allow for the inclusion of covariate information, the generalised Chao estimator does so through regression modelling, resulting in more representative estimates. Since it is a generalised form, in the case where no covariates are provided to the regression model, the generalised Chao estimator is equal to the conventional Chao estimator.

Comparative to the Horvitz-Thompson estimator, this approach has a more relaxed distributional assumption requiring only two consecutive counts to follow the given distribution, rendering it more resilient to outliers. For zero-truncated data, the consecutive counts are typically assumed to be $Y=1$ and $Y=2$, with remaining count values truncated. The resulting truncated likelihood is equal to the standard binomial logistic likelihood, the maximum likelihood estimates of the expected counts as follows
\begin{equation*}
    \hat{\mu}_i=\dfrac{2\hat{q}_i}{1-\hat{q}_i},
\end{equation*}
where $\hat{q}_i$ are the fitted values of the logistic regression model for $i=1,\dots,M$, where $M=(f_1+f_2)$ (please see details in Appendix~\ref{app:truncpoislik}).

Using this maximum likelihood estimate, the generalised Chao population size estimate is given as

\begin{equation*}
\widehat{N}^{(GC)}=n+\sum_{i=1}^{M} \dfrac{f_{i1}+f_{i2}}{\hat{\mu}_i+\hat{\mu}_i^2/2},
\end{equation*}
where $f_{iy}$ is the frequency of exactly $Y=y$ counts in study $i$. 

As with the Horvitz-Thompson estimator, the theoretical formula in (\ref{equ:condvar}) proposed by \citet[][page 314]{van2003point} can be used to find the analytical variance. \cite{bohning2013generalization} approximates this variance for the generalised Chao estimator to be
\begin{equation}
\begin{aligned}
        \widehat{\text{Var}}(\widehat{N}^{(GC)})=&\left(\sum_{i=1}^{f_1+f_2} \nabla G(\hat{\mu}_i|\hat{\boldsymbol{\beta}})\right)^T\text{Cov}(\hat{\boldsymbol{\beta}})\left(\sum_{i=1}^{f_1+f_2} \nabla G(\hat{\mu}_i|\hat{\boldsymbol{\beta}})\right)+ \\
        &\sum_{i=1}^{f_1+f_2}(1-\hat{q}_i)\left(1+\dfrac{\exp(-\hat{\mu}_i)}{\hat{q}_i}\right)^2,
        \end{aligned}
        \label{equ:vargc}
\end{equation}
where
\begin{equation*}
    \nabla G(\hat{\mu}_i|\boldsymbol{\hat{\beta}})=\dfrac{\hat{\mu}_i+\hat{\mu}_i^2}{(\hat{\mu}_i+\hat{\mu}_i^2/2)^2}\times\mathbf{h}(\mathbf{x}_i)^T.
\end{equation*}

\subsection{Generalised Zelterman estimator}\label{sec:genZelt}
As with the conventional Chao's estimator, the conventional Zelterman estimator, developed by \cite{zelterman1988robust}, assumes that only a small range of count values follow the given distribution, improving its resilience to outliers. The expected count for each study is estimated as
\begin{equation*}
    \hat{\mu}=\dfrac{(k+1)f_{k+1}}{f_k}.
\end{equation*}
For zero-truncated data, typically $k=1$ is assumed, since the missing frequency $f_0$ is close to the frequencies $f_1$ and $f_2$. In this case, all counts besides $Y=1$ and $Y=2$ are truncated, leading to the population size estimator $N^{(Z)}=n/(1-\exp(-\hat{\mu}))$. However, the conventional approach also does not allow for covariate information, motivating the truncated maximum likelihood estimate approach of the generalised Zelterman estimator developed by \cite{bohning2009covariate}. As with the generalised Chao estimator, if no covariates are included in the modelling, the generalised Zelterman estimator is equal to the conventional Zelterman estimator.

Using the same binomial logistic likelihood as in Section~\ref{sec:genChao}, the binary outcome probability, $q_i$, is connected via a logit link to the linear predictor from the regression model and the expected count parameter respectively as
\begin{equation*}
    q_i=\dfrac{e_i\exp(\mathbf{h}(\mathbf{x}_i)^T\boldsymbol{\beta})}{1+e_i\exp(\mathbf{h}(\mathbf{x}_i)^T\boldsymbol{\beta})}, \text{ and }
    q_i=\dfrac{\mu_i/2}{1+\mu_i/2}.
\end{equation*}
Therefore, the expected count can be estimated as $\hat{\mu}_i=2e_i\exp(\mathbf{h}(\mathbf{x}_i)^T\boldsymbol{\hat{\beta}})$, for $i=1,\dots,n$.

Using this value of the parameter in the conventional Zelterman estimator, accounting for covariate information, leads to the generalised Zelterman estimator, given formally as

\begin{equation*}
\widehat{N}^{(GZ)}=\sum_{i=1}^{n} \dfrac{1}{1-\exp(-\hat{\mu}_i)},
\end{equation*}
for $i=1,\dots,n$.

The conditioning approach by \citet[][page 314]{van2003point} can also be applied to the generalised Zelterman estimator. Following the work of \cite{bohning2009covariate}, the analytical variance is approximated as
\begin{equation}
\begin{aligned}
    \widehat{\text{Var}}(\widehat{N}^{(GZ)})=&\left(\sum_{i=1}^{n} \nabla G(\hat{\mu}_i|\hat{\boldsymbol{\beta}})\right)^T\text{Cov}(\hat{\boldsymbol{\beta}})\left(\sum_{i=1}^{n} \nabla G(\hat{\mu}_i|\hat{\boldsymbol{\beta}})\right)+ \\
    &\sum_{i=1}^n\dfrac{\exp(-\mu_i)}{(1-\exp(-\mu_i))^2},
\end{aligned}
\label{equ:vargz}
\end{equation}
where
\begin{equation*}
    \nabla G(\hat{\mu}_i|\boldsymbol{\hat{\beta}})=-\dfrac{\exp(\log(\hat{\mu}_i)-\hat{\mu}_i)}{(1-\exp(-\hat{\mu}_i))^2}\times \mathbf{h}(\mathbf{x}_i)^T.
\end{equation*}

\subsection{Application}
Each of the capture-recapture population size estimators given in this section can be applied to the case study data. To account for the covariate information, the various linear predictors given in Table~\ref{tab:lp} in Appendix~\ref{app:tables} are considered, with the best fitting model selected using the BIC.
For the Horvitz-Thompson estimator, the choice of distribution is between the Poisson and negative-binomial distributions, given the nature of count data. However, the generalised Chao and generalised Zelterman estimator do not have a choice of distribution given, with a binomial logistic regression model requiring to be fitted.

\begin{table}[ht]
\caption{Values of the BIC statistic for models under consideration, where the Poisson and negative-binomial distributions model the full data and the binomial distribution models the truncated data. Preferred models are indicated in bold text.  \label{tab:bic}}
\begin{center}
{\small
\begin{tabular}{lrrr}
\toprule
    Distribution & Linear predictor & Log-likelihood & BIC \\ \hline
     & \textbf{1} & \textbf{-23.7} & \textbf{50.7} \\ 
     & 2 & -23.4 & 53.4 \\
Poisson  & 3 & -23.0 & 52.6 \\
     & 4 & -23.0 & 55.9 \\
    \textit{(Full data)} & 5 & -22.7 & 58.6 \\\hline     
     & 1 & -23.7 & 54.0 \\
     & 2 & -23.4 & 56.7 \\
Negative-binomial & 3 & -23.0 & 55.9 \\
      & 4 & -23.0 & 59.2 \\
     \textit{(Full data)} & 5 & -23.7 & 61.9 \\\hline \hline
     & \textbf{1} & \textbf{-7.8} & \textbf{18.6} \\ 
    & 2 & -7.0 & 20.2  \\
Binomial & 3 & -7.8 & 21.6 \\
     & 4 & -7.0 & 23.2 \\
    \textit{(Truncated data)} & 5 & -5.7 & 23.5 \\\bottomrule
\end{tabular}}
\end{center}
\end{table}
Table~\ref{tab:bic} gives the log-likelihood and BIC statistic values for each of the linear predictors and distribution combinations under consideration. The Horvitz-Thompson estimator utilises the entire data available, and hence the Poisson and negative-binomial distributions model with the full data. However, the generalised Chao and generalised Zelterman estimators use the truncated data, containing only counts of $Y=1$ and $Y=2$, so the binomial distribution models using the truncated data. Since the data used for the binomial models differs from the other models, the results cannot be directly compared. Between the models within each distribution, there is little change in the log-likelihood values, and negligible difference between the Poisson and the negative-binomial distributions. Therefore, utilising BIC statistics for model selection, the preferred model to be used in the Horvitz-Thompson estimator is the intercept-only Poisson model, and for the generalised Chao and generalised Zelterman estimators, the intercept-only binomial model is preferred.

Utilising the preferred models in the application of the Horvitz-Thompson, generalised Chao and generalised Zelterman estimators respectively, leads to the estimated total number of studies of $\widehat{N}^{(HT)}=134$, $\widehat{N}^{(GC)}=173$ and $\widehat{N}^{(GZ)}=175$. It is to be expected that the generalised Chao is comparable to, but slightly lower than, the generalised Zelterman estimate given the similarity of the methods and that the generalised Chao is a lower bound estimator. However, they both differ largely from the Horvitz-Thompson estimate which is much smaller as a result of the difference in distributional assumptions and models used. The difference in population size estimates can lead to differing conclusions, and hence the estimator chosen can have an impact on the reliability of any conclusions made. 

As for the uncertainty assessment for the estimators, using (\ref{equ:varht}),(\ref{equ:vargc}) and (\ref{equ:vargz}), the variances for the Horvitz-Thompson, generalised Chao and generalised Zelterman estimators are $\widehat{\text{Var}}(\widehat{N}^{(HT)})=1677$, $\widehat{\text{Var}}(\widehat{N}^{(GC)})=12707$ and $\widehat{\text{Var}}(\widehat{N}^{(GZ)})=13425$ respectively. The $95\%$ confidence interval for each estimator is computed as
\begin{equation*}
    CI=\widehat{N} \pm 1.96 \sqrt{\widehat{\text{Var}}(\widehat{N})},
\end{equation*}
leading to the corresponding confidence intervals $CI^{(HT)}=(51, 214)$, $CI^{(GC)}=(27, 394)$ and $CI^{(GZ)}=(27, 402)$.

The latter two confidence intervals are twice the width of the interval from the Horvitz-Thompson estimator, indicating that there is considerably more uncertainty associated with using the generalised Chao and generalised Zelterman. This increased uncertainty is expected given that the observed number of studies is already small to estimate from, and the generalised Chao and generalised Zelterman estimators truncate the data further, estimating from an even smaller sample size, and the more data available, the less uncertainty there is when estimating. Since the Horvitz-Thompson estimator utilises the entire data available, the uncertainty is reduced, leading to a smaller variance and narrower confidence interval which has a lower limit greater than the lower bound for the total number of studies, making it a more reliable confidence interval to draw conclusions from.

\begin{table}[ht!]
\caption{Simulated data with values of the number of completed suicides, person-years, proportion of women and country of origin of the studies, where the number of completed suicides are sampled from an alternative distribution to be outliers.\label{tab:outliers}}
\begin{center}
{\small \begin{tabular}{lrrrr}
\toprule
Study & Number of & Person-& Proportion & Country \\
 & completed suicides & years & of women & of origin \\
  $i$&$y_i$& $e_i$ & $x_{i1}$ & $x_{i2}$\\
\hline
28. & 14 & 1862 & 0.8371998 & Other \\
29. & 17 & 2410 & 0.8087218 & USA \\
30. & 16 & 1951 & 0.8430489 & Other \\ \bottomrule
\end{tabular}}
\end{center}
\end{table}

However, the question of what happens if the model and distributional assumptions are not met remains, for example, how do outliers affect the results and inference. To address the question of the impact outliers have on the estimators and corresponding conclusions, outliers can be added to the case study data. The observed rates for each study, computed by dividing the number of completed suicides by the person-years, are used to find the lower bound for which rates are classed as outliers.  Formally, the lower bound for a rate to be classified as an outlier is computed as
\begin{equation}
    \lambda^L = Q3 + 3\times IQR,
    \label{equ:outliers}
\end{equation}
where $Q3$ is the third quartile and $IQR$ is the inter-quartile range for the observed rate. To mimic the variability of rates between studies experienced in real life data, the outlier rates are sampled from a uniform distribution with a lower bound of $\lambda^L$ and an upper bound, $\lambda^U$, computed as
\begin{equation}
    \lambda^U=1.2\times\lambda^L.
    \label{equ:outliers2}
\end{equation}
To convert the outlier rates into counts, person-years is required, found by sampling the number of participants in each study from the Poisson distribution and observational period from the log-normal distribution and multiplying the respective values for each study. The sampled person-years multiplied by the outlier rates produces counts of completed suicide which are classified as outliers for the data. With only 27 observed studies, the addition of 3 studies with counts that are outliers leads to a proportion of 10\% of the observed data being outliers, and a proportion of approximately 2\% of the total data. Table~\ref{tab:outliers} displays the values of the number of completed suicides, person-years, proportion of women and country of origin of the 3 additional studies. Utilising the Horvitz-Thompson, generalised Chao and generalised Zelterman estimators respectively, the estimated total number of studies are $\widehat{N}^{(HT)}=479717$, $\widehat{N}^{(GC)}=176$ and $\widehat{N}^{(GZ)}=180$, with corresponding 95\% confidence intervals are $CI^{(HT)}=(30, 47285488)$, $CI^{(GC)}=(30, 397)$ and $CI^{(GZ)}=(30, 411)$.

There is no notable impact from outliers on the generalised Chao and generalised Zelterman estimators, with the estimates after outliers differing only slightly from the number of outliers studies included and the number of studies estimated from the data without outliers combined. However, the number of total studies found using the Horvitz-Thompson estimator is increased significantly after the inclusion of outliers to a number of studies which is inaccurate, with the corresponding confidence interval indicating a large quantity of uncertainty with its width.

As the results differ by such a large margin, it is important to know which of the estimators produces the most reliable results for the given data and hence the best conclusions. This importance motivates the use of a simulation study to compare the performance of the estimators, testing various data scenarios and sizes to assess which estimator is most applicable to use and when.

\section{Simulation} \label{sec:simstudy}
To create a data set where for each study $i=1,\dots,N$, the values for counts, person-years and covariates are simulated to reflect the values in the case study, certain parameters require defining. For simulating person-years, the mean number of participants per study, $\Bar{t}$,  logarithm of the mean $\gamma$ and standard deviation $\sigma$ of the observational period are required. Using the sampled person-years, and a constant rate of event $\lambda^C$, the count values can be simulated. As the covariates, $\alpha$ and $\beta$ are shape parameters for a beta distribution to simulate $x_1$, and the probability of success for $x_2$ is given by $\rho$.

Using the predefined parameters, the number of participants in each study is sampled from the Poisson distribution, $t_i~\sim~Poisson(\Bar{t})$, and the observational period for each study is sampled from the log-normal distribution, $O_i~\sim~lognormal(\gamma, \sigma)$, leading to the sampled exposure variable of person-years, $\tau_i=t_i\times O_i$. The count of events for each study is then sampled from the binomial distribution as $X_i~\sim~binomial(\tau_i, \lambda^C)$. To include covariate information in the simulation study, the numeric variable for the proportions is sampled from the beta distribution, $x_1~\sim~beta(\alpha, \beta)$, and the binary variable for the country of origin is sampled from the Bernoulli distribution, $x_2~\sim~Bernoulli(\rho)$.

The sampling process is repeated $S$ times, creating a zero-truncated data set for each iteration though removing all studies which have a count of $Y_i=0$. To this data, the capture-recapture estimators and respective analytical variances discussed in Section~\ref{sec:estimation} can be computed, producing population size estimates and respective confidence intervals. 

Performance of the estimators is assessed using the following measures.
\begin{itemize}
    \itemsep0mm
    \item[-] Accuracy: 
    \begin{equation*}
        median(|\boldsymbol{\widehat{N}}-N|),
    \end{equation*}
    where $\boldsymbol{\widehat{N}}=(\widehat{N}_1,\dots,\widehat{N}_S)$ are the estimated population sizes from each iteration of the simulation study and $N$ is the true population size.
    \item[-] Precision: 
    \begin{equation*}
        median(\boldsymbol{CI}_U-\boldsymbol{CI}_L),
    \end{equation*}
    where $\boldsymbol{CI}_L=CI_{L,1},\dots,CI_{L,S}$ and $\boldsymbol{CI}_U=CI_{U,1},\dots,CI_{U,S}$ respectively are the lower and upper limits of the 95\% confidence intervals for the estimated population size for each iteration of the simulation study.
    \item[-] Coverage: 
    \begin{equation*}
    \frac{1}{S}\sum_{s=1}^S J_s \times 100\%,
    \end{equation*}
    where for $s=1,\dots,S$, $J_s$ is an indicator variable defined as
    \begin{equation*}
        J_s=\begin{cases}
            1 & \text{ if }CI_{L,s} \leq N \leq CI_{U,s} \\
    0 & \text{ otherwise}.
    
        \end{cases}
    \end{equation*}
    \item[-] Robustness: Defined as the resilience of the estimator to outliers. 
    \begin{itemize}
    \itemsep0mm
        \item[-] In the simulation study, it is measured through comparing the values for accuracy, precision and coverage for data without outliers to values for data with outliers. To simulate the outlier counts, the person-years are multiplied by an outlier rate sampled from the uniform distribution, $\lambda_i^O~\sim~uniform(\lambda^L, \lambda^U)$, where the boundary values are chosen by an approximation of the results from (\ref{equ:outliers}) and (\ref{equ:outliers2}) applied to the data being replicated. Given that the order of the studies does not impact the modelling results, the defined proportion of outliers are included at the end of the data.
    \end{itemize} 
\end{itemize}

\section{Results} \label{sec:results}

\begin{figure}[htbp!]
    \centering
    \includegraphics[width=0.8\linewidth]{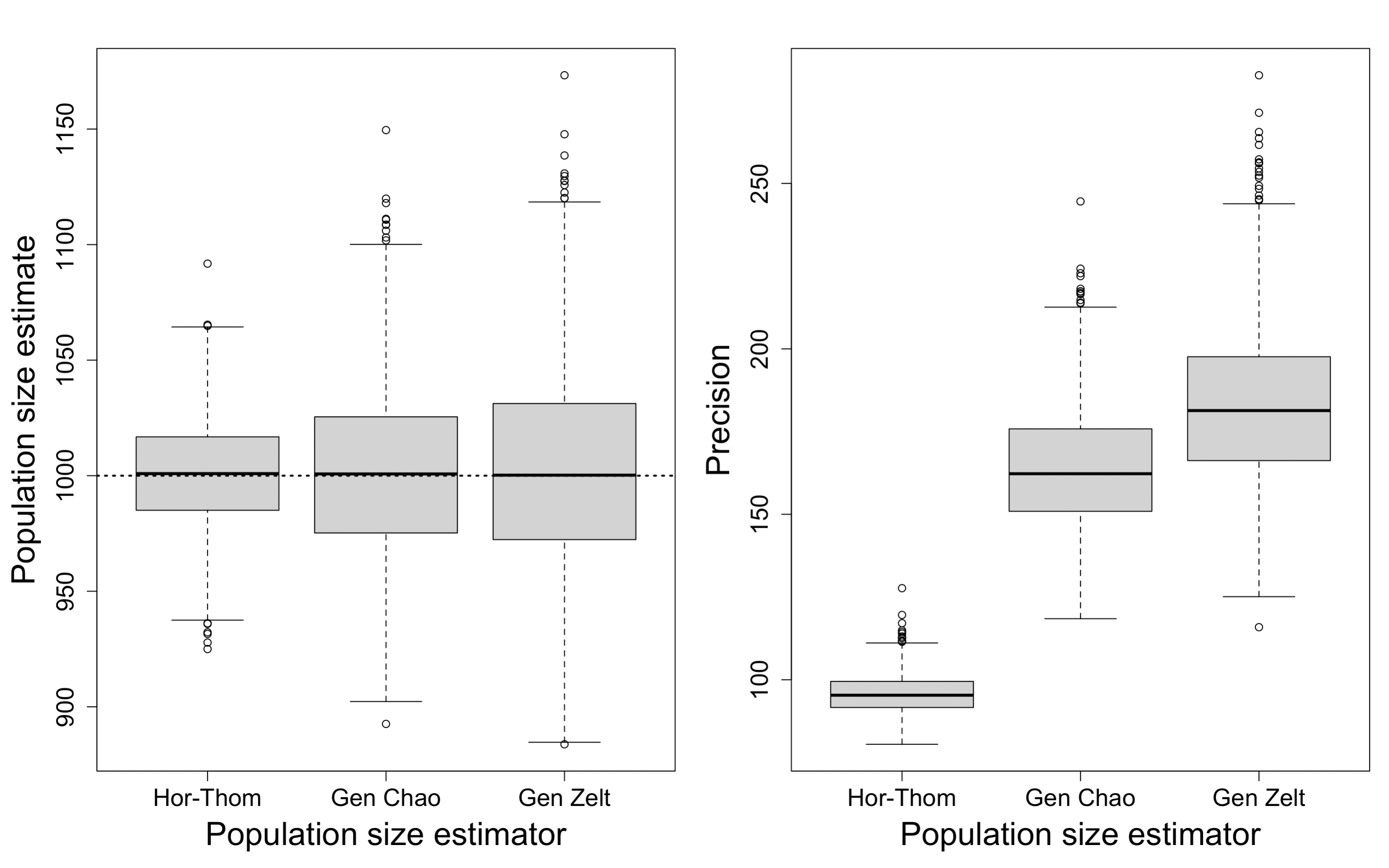}
    \caption{Box plots showing the values of the population size estimates (left) and the values for the precision of the confidence intervals (right) for the Horvitz-Thompson, generalised Chao and generalised Zelterman estimators when there are no outliers in the data and the dashed line represents the true population size of $N=1000$.}
    \label{fig:plot0percN1000}
\end{figure}

\begin{figure}[ht!]
    \centering
    \includegraphics[width=0.8\linewidth]{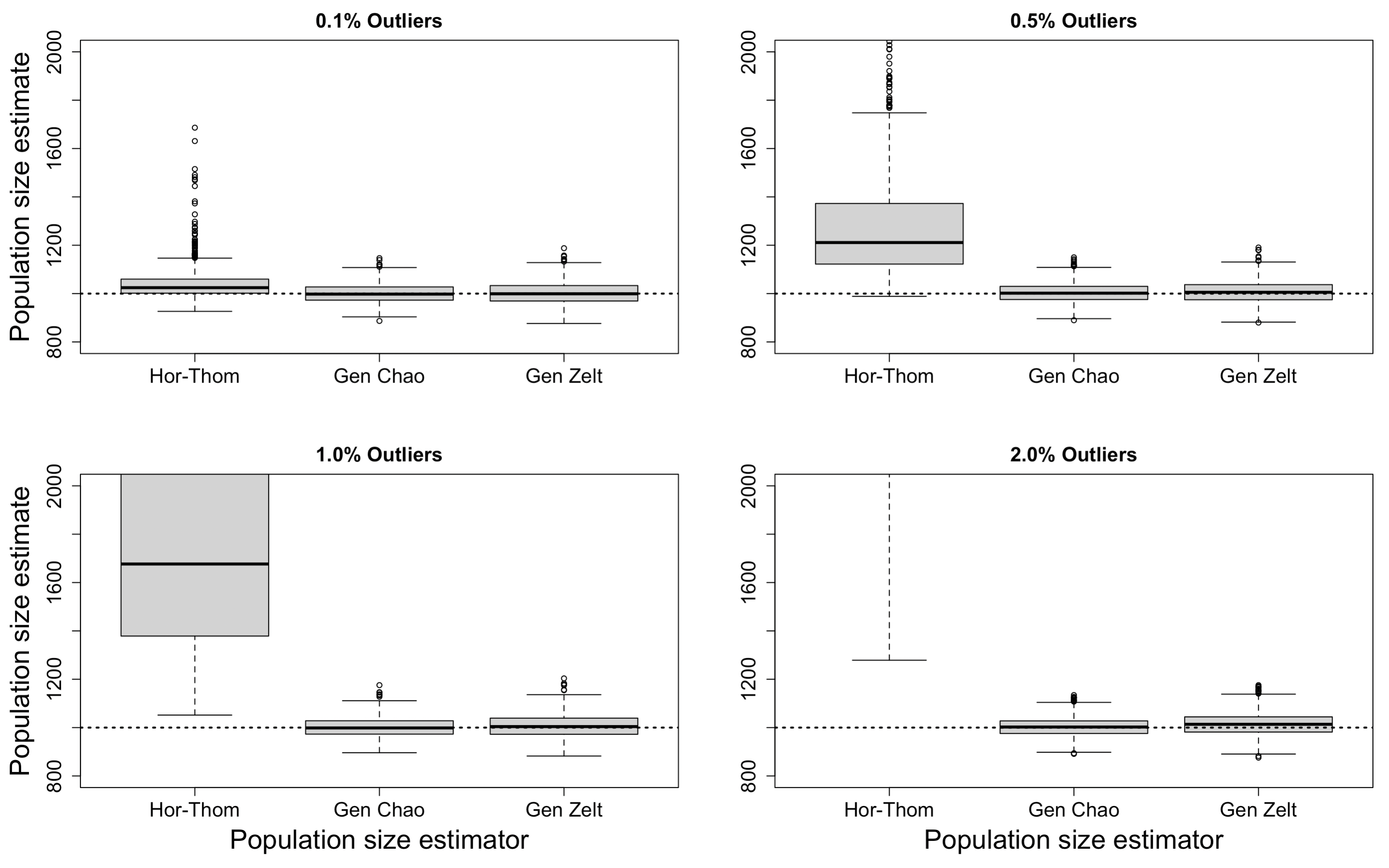}
    \caption{Box plots showing the values of the population size estimates for the Horvitz-Thompson, generalised Chao and generalised Zelterman estimators and varying proportions of outliers. The dashed line represents the true population size of $N=1000$.}
    \label{fig:N1000outliersaccuracy}
\end{figure}

\begin{figure}[ht!]
    \centering
    \includegraphics[width=0.8\linewidth]{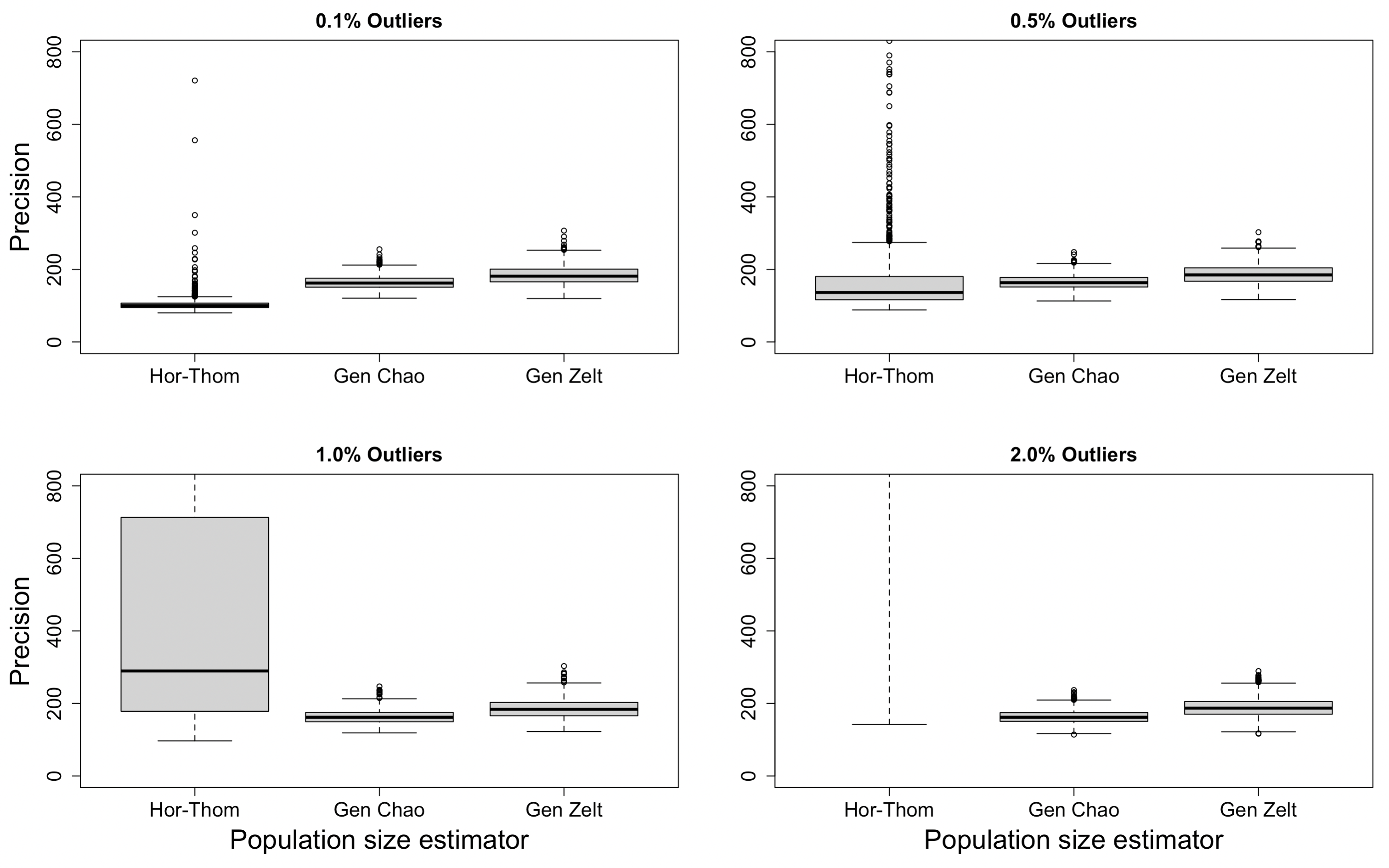}
    \caption{Box plots showing the values of the precision from the 95\% confidence intervals for the Horvitz-Thompson, generalised Chao and generalised Zelterman estimators and varying proportions of outliers when the true population size is $N=1000$.}
    \label{fig:N1000outliersprecision}
\end{figure}

Table~\ref{tab:CRE_N1000sim} displays the values of accuracy, precision and coverage for the Horvitz-Thompson, generalised Chao and generalised Zelterman estimators, and total number of studies is $N=1000$. The performance measures are given for proportions of outliers varying from $0\%$ to $2\%$ to also assess robustness of each estimator. When the counts follow the distributional assumptions perfectly, the Horvitz-Thompson estimator is both the most accurate and the most precise, illustrated in Figure~\ref{fig:plot0percN1000} with the smallest inter-quartile and total ranges for both measures occurring for the Horvitz-Thompson estimator, and the corresponding plot for precision being closer to zero than the alternative estimators. Whilst the generalised Chao has the highest coverage, for all the estimators coverage is desirable at at least $95\%$, with negligible difference. However, as outliers are introduced to the data, the preference for the Horvitz-Thompson estimator becomes less obvious. Up to $0.5\%$ of the counts being outliers, the Horvitz-Thompson estimator has the best performance for precision, however, once more outliers are introduced to the data, precision is dramatically reduced. Additionally, with as little as $0.1\%$ outliers, the accuracy of the Horvitz-Thompson estimator is impacted, and coverage is significantly decreased, with only $70\%$ of the confidence intervals containing the true value. As the proportion of outliers increases, the performance of the Horvitz-Thompson estimator worsens, with estimates getting further from the true value and confidence intervals getting wider from increased uncertainty. It appears that past a certain proportion of outliers, the coverage begins to improve, with coverage having an increase of $50\%$ between $1\%$ and $2\%$ outliers, however, this is due to the width of the intervals growing, increasing the changes of the interval to contain the true value. Changes in the accuracy and precision of the population size estimates respectively are illustrated in Figures~\ref{fig:N1000outliersaccuracy} and ~\ref{fig:N1000outliersprecision}, where the dispersion of the Horvitz-Thompson values increases as the proportion of outliers increases, and the median values grow farther from either the true population size or a reasonable width of confidence interval.
\begin{table}[htbp!]
    \centering
    \caption{Values for the reliability measures of accuracy, precision and coverage for the capture-recapture population size estimators of Horvitz-Thompson, generalised Chao and generalised Zelterman, where $S=1000$, $N=1000$, $\Bar{t}=900$, $\lambda^C=0.0004$, $\lambda^L=0.007$, $\lambda^U=0.009$, $\gamma=1.5$, $\sigma=0.8$, $\alpha=36$, $\beta=8.5$ and $\rho=0.4$ for various proportions of outliers.} \label{tab:CRE_N1000sim}
    {\small
    \begin{tabular}{ccccccc}
         \toprule
          &  & \multicolumn{5}{c}{Proportion of Outliers} \\ \cline{3-7}
         Measure & Estimator & 0.0\% & 0.1\% & 0.5\% & 1.0\% & 2.0\% \\ \hline
          & Horvitz-Thompson & 16 & 30 & 211 & 677 & 2.1e+06 \\
         Accuracy & Generalised Chao & 25 & 27 & 27 & 27 & 26 \\
          & Generalised Zelterman & 29 & 32 & 31 & 32 & 32 \\ \hline
          & Horvitz-Thompson & 95 & 100 & 136 & 290 & 6.7e+07 \\
         Precision & Generalised Chao & 162 & 162 & 163 & 162 & 162 \\
          & Generalised Zelterman & 181 & 181 & 185 & 184 & 187 \\ \hline
          & Horvitz-Thompson & 95.5\% & 69.6\% & 7.0\% & 7.4\% & 60.6\% \\
         Coverage & Generalised Chao & 96.4\% & 96.0\% & 96.4\% & 96.7\% & 95.6\% \\
          & Generalised Zelterman & 95.7\% & 94.7\% & 95.8\% & 96.7\% & 94.8\% \\ 
          \bottomrule
    \end{tabular}}
\end{table}

\begin{table}[ht]
    \centering
    \caption{Values for the reliability measures of accuracy, precision and coverage for the capture-recapture population size estimators of Horvitz-Thompson, generalised Chao and generalised Zelterman, where $S=1000$, $N=500$, $\Bar{t}=900$, $\lambda^C=0.0004$, $\lambda^L=0.0004$, $\lambda^U=0.0004$, $\gamma=1.5$, $\sigma=0.8$, $\alpha=36$, $\beta=8.5$ and $\rho=0.4$ for various proportions of outliers. Number of outliers required to be integers so values for the proportion of 0.1\% outliers are not given.} \label{tab:CRE_N500sim}
    {\small
    \begin{tabular}{ccccccc}
         \toprule
          &  & \multicolumn{5}{c}{Proportion of Outliers} \\ \cline{3-7}
         Measure & Estimator & 0.0\% & 0.1\% & 0.5\% & 1.0\% & 2.0\% \\ \hline
          & Horvitz-Thompson & 11 & - & 62 & 293 & 4797 \\
         Accuracy & Generalised Chao & 19 & - & 19 & 18 & 19 \\
          & Generalised Zelterman & 21 & - & 22 & 21 & 22 \\ \hline
          & Horvitz-Thompson & 67 & - & 83 & 181 & 10096 \\
         Precision & Generalised Chao & 116 & - & 116 & 115 & 113 \\
          & Generalised Zelterman & 130 & - & 131 & 130 & 130 \\ \hline
          & Horvitz-Thompson & 94.8\% & - & 34.8\% & 14.5\% & 51.1\% \\
         Coverage & Generalised Chao & 96.9\% & - & 96.3\% & 96.7\% & 95.5\% \\
          & Generalised Zelterman & 94.6\% & - & 96.7\% & 95.7\% & 95.0\% \\ 
          \bottomrule
    \end{tabular}}
\end{table}

For completeness, Table~\ref{tab:CRE_N500sim} demonstrates the effect of outliers on the performance of the estimators when the total number of studies differs, specifically when $N=500$. As with when $N=1000$, when all data follows the distribution, the Horvitz-Thompson estimator performs the best, but the preference changes to the generalised Chao and generalised Zelterman estimators once outliers are introduced. 

Throughout both tables, there is little difference between the performance for the generalised Chao and generalised Zelterman estimators, for both varying total sizes of data and proportions of outliers. However, the generalised Chao estimator is consistently more accurate and almost always has slightly higher coverage. The only variation is as a result of data sizes, where for larger data sets, the generalised Chao estimator is more precise, and the generalised Zelterman estimator more precise for smaller data sets. 

 The values in Tables~\ref{tab:CRE_N1000sim} and \ref{tab:CRE_N500sim} suggest that it is the number of outliers rather than the proportion of outliers that impact the Horvitz-Thompson estimator's performance. For each proportion of outliers included respectively, comparing the performance of the Horvitz-Thompson estimator for N=1000 and N=500 indicate that the larger study size of N=1000 impacts the estimator more, with a reduction in accuracy, coverage and precision. However, if the number of outliers is used as the comparison measure, rather than the proportion of outliers, the performance of the estimator is much more comparable. For example, when 5 outliers are included in the data, the proportion of outliers is 0.5\% for N=1000 and 1.0\% for N=500. For these proportions, the values of accuracy, precision and coverage respectively are more comparable and differ less from each other with the different population sizes than when a proportion of 0.5\% outliers is used for N=500. Given these results are replicated for the other proportions simulated, it is important for data to follow the distributional assumptions for the Horvitz-Thompson estimator to be used, as even for a very large population size, a very small number of outliers impact its performance. 

Overall, the Horvitz-Thompson estimator performs better than the alternative estimators when the data follows the distributional assumption given. However, this is more often than not not the case as a result of unpredictability within real life populations. Therefore, assumptions are not always met and in the presence of outliers, the generalised Chao and generalised Zelterman estimators are the preferred estimator, given they are more robust.

\section{Discussion}
This paper explores the performance of different capture-recapture population size estimators when dealing with zero-truncated meta-analytic count data through utilising a simulation study. A benefit of this approach is the flexibility enabled when creating the data sets, allowing for different data scenarios to be applied and covariate information included to test the performance of the estimators more thoroughly. 

The results from the simulation study indicate a preference for the Horvitz-Thompson estimator only if the data does not contain outliers. Given that within real life data, it is a common occurrence for outliers to be included, and even if it is only a small proportion, the Horvitz-Thompson estimator is not the most reliable. Between the generalised Chao and generalised Zelterman estimators, there is very little difference in performance, with the reliability measures unaffected by outliers, demonstrated by the consistent desirable coverage in addition to appropriate accuracy and precision irrespective of the proportion of outliers. The negligible difference in performance means that either estimator is appropriate and would return reliable results, but specifically for larger data, the generalised Chao is favoured, and the generalised Zelterman favoured for smaller data. 

For future work, additional data structures could be explored, such as data with different covariate variable types or sampling distributions assumed, to examine the estimators' performance in a wider range of scenarios. It could also prove beneficial to explore their performance using alternative confidence interval construction methods like the bootstrap algorithm and the percentile method, given that the analytical approach taken in this paper does not produce appropriate intervals for the small number of studies from the case study used with the generalised Chao and generalised Zelterman estimators. Lastly, the estimators discussed in this paper are not the only capture-recapture estimators available so the performance of a wider range of estimators could be explored. For example, the Turing estimator \cite{good1953population} and conventional Chao and Zelterman estimators discussed in Section~\ref{sec:estimation}, whilst not appropriate for the simulation study in this paper given that they do not allow for covariates, for different data scenarios where covariates and exposure variables are not included, a comparison of performance could lead to more reliable estimator selection. 

\bibliographystyle{bibstyle}

\appendix
\renewcommand{\thetable}{\thesection.\arabic{table}}
\setcounter{table}{0}

\section{Appendix}

\subsection{Tables}\label{app:tables}
\begin{table}[ht!]
\caption{Linear predictors for models under consideration. \label{tab:lp}}
\begin{center}
{\small
\begin{tabular}{rrrrr}
\toprule
Linear & Proportion & Country & Interaction &  \\ 
predictor & of women, $x_1$ & of origin, $x_2$ & $x_1x_2$ & $\bh(\bx)$\\ \hline 
1 & No & No & No & $\bh_1(\bx) = 1$\\
2 & Yes & No & No & $\bh_2(\bx) = (1,x_1)^T$\\
3 & No & Yes & No & $\bh_3(\bx) = (1,x_2)^T$ \\
4 & Yes & Yes & No & $\bh_4(\bx) = (1,x_1,x_2)^T$ \\
5 & Yes & Yes & Yes& $\bh_5(\bx) = (1,x_1,x_2,x_1x_2)^T$ \\ \bottomrule
\end{tabular}}
\end{center}
\end{table}

\begin{table}[ht]
\caption{Meta-analytic data from \cite{peterhansel2013risk}, numbered and ordered by decreasing size of person-years. The table includes the number of person-years, the proportion of women, the country of origin and the number of completed suicides for each study. The proportion of women for 24. Smith 2004 is unknown but is imputed to be 0.823. The country of origin for 21. Kral 1993 is reported as ``USA/Sweden" but changed to USA for model fitting.\label{tab:datatable}}
\begin{center}
{\footnotesize \begin{tabular}{lrrrr}
\toprule
Study & Number of & Person-& Proportion & Country \\
 & completed suicides & years & of women & of origin \\
  $i$&$y_i$& $e_i$ & $x_{i1}$ & $x_{i2}$\\
\hline
1. Adams 2007 & 21 & 77602 & 0.860 & USA\\ 
2. Marceau 2007 & 6 & 10388 & 0.720 & Canada\\ 
3. Marsk 2010 & 4 & 8877 & 0.000 & Sweden\\ 
4. Pories 1995 & 3 & 8316 & 0.832 & USA\\ 
5. Carelli 2010 & 1 & 6057 & 0.684 & USA\\ 
6. Busetto 2007 & 1 & 4598 & 0.753 & Italy\\
7. Smith 1995 & 2 & 3882 & 0.889 & USA\\
8. Peeters 2007 & 1 & 3478 & 0.770 & Australia\\
9. Christou 2006 & 2 & 2599 & 0.820 & Canada\\
10. G\"unther 2006 & 1 & 2244 & 0.837 & Germany\\
11. Capella 1996 & 3 & 2237 & 0.822 & USA\\
12. Suter 2011 & 3 & 2152 & 0.744 & Switzerland\\
13. Suter 2006 & 1 & 1639 & 0.865 & Switzerland\\
14. Van de Weijgert 1999 & 1 & 1634 & 0.870 & Netherlands\\
15. Cadi\`{e}re 2011 & 1 & 1362 & 0.834 & Belgium\\
16. Mitchell 2001 & 1 & 1121 & 0.847 & USA\\
17. Himpens 2011 & 1 & 1066 & 0.902 & Belgium\\
18. N\"aslund 1994 & 2 & 799 & 0.812 & Sweden\\
19. Forsell 1999 & 1 & 761 & 0.761 & Sweden\\
20. Powers 1997 & 1 & 747 & 0.847 & USA\\
21. Kral 1993 & 1 & 477 & 0.812 & \textit{USA}\\
22. N\"aslund 1995 & 1 & 457 & 0.592 & Sweden\\
23. Powers 1992 & 1 & 395 & 0.850 & USA\\ 
24. Smith 2004 & 1 & 354 & \emph{0.823} & USA\\ 
25. Nocca 2008 & 1 & 228 & 0.677 & France\\ 
26. Svenheden 1997 & 1 & 166 & 0.791 & Sweden\\ 
27. Pekkarinen 1994 & 1 & 146 & 0.704 & Finland \\ \bottomrule
\end{tabular}}
\end{center}
\end{table}

\subsection{The truncated Poisson likelihood}\label{app:truncpoislik}
This is largely following \cite{bohning2013generalization}.
For $i=1,\dots,M$ where $M=f_1+f_2$ is the number of studies with one or two events, let
\begin{equation*}
    \mu_i=e_i\exp(\mathbf{h}^{T}(\mathbf{x}_i)\boldsymbol{\beta})=e_i\exp(\beta_0+\mathbf{h}^{*T}(\mathbf{x}_i)\boldsymbol{\beta}^*)
\end{equation*}
where $\boldsymbol{\beta}=(\beta_0, \boldsymbol{\beta}^*)$ and $\mathbf{h}^*$ is equal to $\mathbf{h}$ without the intercept. 

The Poisson likelihood truncated for all counts except ones and twos is 
\begin{equation*}
\begin{aligned}
     \prod_{i=1}^M&\left(\dfrac{1}{1+\mu_i/2}\right)^{f_{i1}}\left(\dfrac{\mu_i/2}{1+\mu_i/2}\right)^{f_{i2}}, \\
    =\prod_{i=1}^M&\left(\dfrac{1}{1+e_i\exp(\beta_0+\mathbf{h}^{*T}(\mathbf{x}_i)\boldsymbol{\beta}^*)}\right)^{f_{i1}}\left(\dfrac{e_i\exp(\beta_0+\mathbf{h}^{*T}(\mathbf{x}_i)\boldsymbol{\beta}^*)}{1+e_i\exp(\beta_0+\mathbf{h}^{*T}(\mathbf{x}_i)\boldsymbol{\beta}^*)}\right)^{f_{i2}}, \\
    = \prod_{i=1}^M &(1-q_i)^{f_{i1}}q_i^{f_{i2}},
\end{aligned}
\end{equation*}
hence, $q_i=\dfrac{1}{1+\mu_i/2}$.

This is a conventional binomial logistic likelihood and can be further written as 
\begin{equation*}
    \prod_{i=1}^M \left(\dfrac{1}{1+e_i\exp(\beta_0^{'}+\mathbf{h}^{*T}(\mathbf{x}_i)\boldsymbol{\beta}^*)}\right)^{f_{i1}}\left(\dfrac{e_i\exp(\beta_0^{'}+\mathbf{h}^{*T}(\mathbf{x}_i)\boldsymbol{\beta}^*)}{1+e_i\exp(\beta_0^{'}+\mathbf{h}^{*T}(\mathbf{x}_i)\boldsymbol{\beta}^*)}\right)^{f_{i2}},
\end{equation*}
with $\beta_0^{'}=\log(1/2)+\beta_0$

Once the binomial logistic likelihood has been fitted one can compute 
\begin{equation*}
\hat{\mu}_i=2\frac{\hat{q}_i}{1-\hat{q}_i}=2e_i\exp(\hat{\beta}_0^{'}+\mathbf{h}^{*T}(\mathbf{x}_i)\boldsymbol{\hat{\beta}}^*).
\end{equation*}    
Note that $f_{i1}+f_{i2}=1$ in our case as each study $i$ has either a count of one or a count of two, given it is a truncated study where all counts are truncated except ones and twos.

\end{document}